\documentclass[dvips]{arxstspdf}
\usepackage{flushend}
\usepackage{stfloats}

\volume{26}
\issue{3}
\pubyear{2011}
\firstpage{326}
\lastpage{328}
\doi{10.1214/11-STS352D}
\referstodoi{10.1214/11-STS352}

\begin{document}
\begin{frontmatter}

\title{Discussion of ``Is Bayes Posterior just Quick and Dirty
Confidence?'' by D.~A.~S.~Fraser}
\runtitle{Discussion}
\pdftitle{Discussion of Is Bayes Posterior just Quick and Dirty
Confidence? by D. A. S. Fraser}

\begin{aug}
\author[a]{\fnms{Tong} \snm{Zhang}\ead[label=e1]{tzhang@stat.rutgers.edu}}
\runauthor{T. Zhang}

\affiliation{Rutgers University}

\address[a]{Tong Zhang is Professor, Statistics Department,
Rutgers University, Piscataway, New Jersey 08816, USA \printead{e1}.}

\end{aug}



\end{frontmatter}

\section{Confidence Region Estimation}

The author has written an interesting article on
the relationship of confidence distribution and Baye\-sian posterior distribution.
Confidence distribution has its origin from Fisher's fiducial
distribution, and
in this discussion we refer to it simply as the ``confidence
distribution approach.''
It allows frequentists to assign confidence intervals (or, more
generally, confidence regions) to
the outcome of estimation procedures.

The idea can be simply described as follows. Consider a statistical
model with a family
of distributions $p_\theta(y)$, where $y$ is the observation and $\theta
$ is the model parameter.
We assume that the observed $y$ is generated according to a true
parameter $\theta_*$ which is unknown
to the statistician.
If we can find a real-valued quantity
$U(y;\theta)$ that depends on $\theta$ and $y$ such that for all $\theta$,
when $y$ is generated from $p_\theta(y)$, $U(y;\theta)$ is uniformly
distributed in $(0,1)$,
then we can estimate the confidence interval of $\theta$ given an
observation $y$
as the set $I_{\alpha,\beta}(y)=\{\theta\dvtx  U(y;\theta) \in(\alpha,\beta
)\}$ for some $0 \leq\alpha\leq\beta\leq1$.
An interpretation of this confidence region is that no matter what is
the true underlying $\theta_*$ that generates
$y$, the region $I_{\alpha,\beta}(y)$ contains the true parameter
$\theta_*$ with probability $\beta-\alpha$
(when $y$ is generated according to $\theta_*$).

Indeed, the above interpretation is a very natural definition of
confidence region in the frequentist setting.
It does not assume that $\theta_*$ is generated according to any prior,
and the interpretation holds universally
true for all possible $\theta_*$ in the model.
This interpretation can be compared to a confidence region from the
Bayesian posterior calculation
that assumes that $\theta_*$ is generated according to a specific prior
which has to be known to the statistician.
If the statistician chooses the wrong prior, then the confidence region
calculated from the Bayesian approach will be
incorrect in that it may not contain the true parameter $\theta_*$ with
the correct probability.\looseness=1

The paper takes this interpretation of confidence region, and goes on
to provide several examples showing that
the Bayesian approach does not lead to correct confidence estimates for
all $\theta_*$.
The author then argued that the confidence distribution approach is the
more ``correct'' method for
obtaining confidence intervals and the Bayesian approach is just a
quick and dirty approximation.

One question that needs to be addressed in the confidence distribution
approach is
how to construct a statistics $U(y_0;\theta)$ with the desired
property. The~author considered
the quantity $U(y_0;\theta)=\int_{y \leq y_0} p_\theta(y) \,d y$, which
is well-defined if
the observation $y$ is a real-valued number.
This corresponds to the proposal in Fisher's fiducial distribution.
The idea of fiducial distribution received a number of discussions
throughout the years, and is known to be
adequate for unconstrained location families (for which the fiducial
confidence distribution matches the Bayesian confidence
distribution using a flat prior). However, the general concept is
controversial, and largely regarded as a major blunder by Fisher.

In this discussion article we will explain why the idea of confidence
distribution with
\[
U(y_0;\theta)=\int_{y \leq y_0} p_\theta(y)\, d y
\]
has not received more attention for general statistical estimation
problems, although it does give confidence
region estimates that fit the frequentist intuition.

\section{Suboptimality}

The purpose of confidence distribution is to provide a confidence
region that is consistent with
the frequentist definition.
However, one flaw of this approach is that the result it produces may
not be optimal.
While this issue was pointed out in the article, it was not explicitly
discussed.
In my opinion, this is the main reason why the idea of confidence
distribution hasn't become
more popular in statistics. Therefore, this section provides a more
detailed discussion on this issue.

To understand this point, we shall first consider a simple illustration.
Let $U(y,\theta)$ be a uniform random variable in $(0,1)$ that is
independent of $y$ and~$\theta$.
By definition, given any $\theta_*$, the confidence region $I_{\alpha
,\beta}(y)=\{\theta\dvtx  U(y;\theta) \in(\alpha,\beta)\}$
contains $\theta_*$ with probability $\beta-\alpha$. Since this applies
to the parameter that generates~$y$,
the confidence region obtained this way is consistent with the
frequentist intuition of what
a confidence region should mean. However, this estimate is not useful
statistically
because the method just randomly guesses
either the entire domain of $\theta$ when $U \in(\alpha,\beta)$ or the
empty region
otherwise; the decision does not even depend on $y$.

While the above example is extreme, it does show that a confidence
region merely consistent with
the frequentist semantics is not necessarily a useful estimate.
Statistically, this is because
the confidence region obtained is suboptimal.
In fact, this claim also applies to the confidence distribution
approach this article considers.
More specifically, for nonlinear problems that this paper focused on,
the method can produce
confidence regions that are quite suboptimal.
By ``optimal'' (or even ``good''), we mean that the confidence region a
method produces should be small
by some measure.
In particular, if another method provides confidence regions that also
fit in the frequentist semantics but
is no larger on average for all $\theta$ and smaller for some $\theta$,
then it can be regarded as a better method. This corresponds to the
notion of admissibility in decision theory.

Consider the following simple nonlinear location estimation model:
$y$ is generated either from $N(0,\sigma_0^2)$ when $\theta=0$,
or from $N(1,\sigma_1^2)$ when $\theta=1$. There are only two possible
positions $\theta=0$ or $\theta=1$
for the unknown location parameter $\theta$, and
we assume that the variance parameters $\sigma_0^2$ and $\sigma_1^2$
are known quantities that are not necessarily equal.
Note that the restriction of $\theta$ to two positions is only for
simplicity, which is not critical for our illustration---we can
extend the example to allow all locations in $R$.

For this example, the confidence distribution approach gives the
following $U(y_0,\theta)$:
\[
U(y_0,\theta) =
\cases{
\Phi(y_0/\sigma_0), & $\theta=0,$ \vspace*{2pt}\cr
\Phi\bigl((y_0-1)/\sigma_1\bigr), & $\theta=1 ,$}
\]
where $\Phi(z)$ denotes the cdf of the standard Gaussian $N(0,1)$.

Let's consider the confidence region $I_{\delta,1-\delta}(y)$ for some
$\delta\in(0,0.25)$, which we simplify as
$I(y)$.
By definition, the estimated confidence region $I(y)$ contains the
position $\theta=0$ if and only if $y \in\Omega_0$
with $\Omega_0= (\sigma_0 \Phi^{-1}(\delta), -\sigma_0 \Phi^{-1}(\delta
))$, and $I(y)$ contains the position $\theta=1$ if and only if $y \in
\Omega_1$ with $\Omega_1= (1 + \sigma_1 \Phi^{-1}(\delta), 1- \sigma_1
\Phi^{-1}(\delta))$.
For convenience, we also define
\begin{eqnarray*}
\mu_0&=& P(y \in\Omega_1|\theta=0)\\
&=& \int_{1 + \sigma_1 \Phi^{-1}(\delta
)}^{1-\sigma_1 \Phi^{-1}(\delta)} \frac{1}{\sqrt{2\pi}\sigma_0} \exp
\biggl(- \frac{y^2}{2 \sigma_0^2}\biggr) \,d y .
\end{eqnarray*}
In order to show that the confidence distribution approach is
suboptimal, we can, for simplicity,
consider the case $\sigma_0 \gg1$ and $\sigma_1 \ll1$,
so that $1 - \sigma_1 \Phi^{-1}(\delta) < -\sigma_0 \Phi^{-1}(\delta)$ and
$\mu_0 < 2 \delta$.
The first condition implies that $\Omega_1 \subset\Omega_0$.
Therefore, when the parameter $\theta=1$, with probability $1-P(y \in
\Omega_1|\theta=1)=1-2\delta$ over $y \sim N(1,\sigma_1^2)$, we have $y
\in\Omega_1$ and, thus, $|I(y)|=2$ [i.e., $I(y)$ contains both
$\theta=0$ and $\theta=1$].
Therefore, we have (note that we have assumed that $\delta<0.25$)
%
\begin{equation}
E_{y|\theta=1} |I(y)| > 2 (1-2\delta) > 1 . \label{eq:conf-size-theta-1}
\end{equation}
Moreover, we have
\begin{eqnarray*}
E_{y|\theta=0} |I(y)| &=& P(y \in\Omega_0|\theta=0) + P(y \in\Omega
_1|\theta=0)
\\&=& 1-2\delta+ \mu_0 .
\end{eqnarray*}

Now we would like to construct a better confidence region estimator by
using the condition (which we made earlier) that
$P(y \in\Omega_1|\theta=0) = \mu_0 < 2 \delta$.
Therefore, we can pick $\Omega_0'$ such that $\Omega_0' \cap\Omega_1
=\varnothing$
and $P(y \in\Omega_0'|\theta=0)=1-2\delta$.
This means that we can choose the following confidence region estimate $I'(y)$:
$I'(y)$ contains the position $\theta=0$
if and only if $y \in\Omega_0'$ and $I'(y)$ contains the position
$\theta=1$ if and only if $y \in\Omega_1$.
This estimate obeys the frequentist definition because $P(\theta\in
I'(y)|\theta)=1-2\delta$ both when
$\theta=0$ and $\theta=1$. Moreover, we have
\[
E_{y|\theta=0} |I'(y)| = 1-2\delta+ \mu_0 ,\quad
E_{y|\theta=1} |I'(y)| \leq1 .
\]
The second inequality is due to the fact that $|I'(y)| \leq1$ for all
$y$ because $\Omega_0' \cap\Omega_1=\varnothing$.
In comparison to~(\ref{eq:conf-size-theta-1}), we know that
when $\theta=1$, the confidence distribution approach gives a
confidence region $I(y)$ with a~larger average size.
This means that for this simple problem, the confidence distribution
approach gives a suboptimal estimate
of confidence region $I(y)$ that is dominated by a better method $I'(y)$.
The difference can be significant when $\delta\approx0$.

\section{Conclusion}

The confidence distribution approach is a rather general method to
obtain confidence regions for parameter
estimation problems consistent with the frequentist semantics. The
method can also be easily generalized to
the multivariate situation where $y$ is a vector instead of a real number.
Nevertheless, the confidence region it estimates can be
rather suboptimal in the sense that the region obtained by this method
can be significantly larger than what can be
done with more sophisticated methods.
Although we have only illustrated this phenomenon with a relatively
simple example, the conclusion holds
more generally.

At the root of this suboptimality, we note that
whether a model parameter $\theta_0$ belongs to the confidence region
obtained by the confidence distribution approach
only depends on the distribution $p(y|\theta=\theta_0)$ at the
parameter $\theta_0$ itself,
without considering the alternative models at $\theta\neq\theta_0$.
This unnatural behavior is what causes its suboptimality for general
nonlinear models.
For example, in order to achieve good performance
for the simple two-position location estimation example given in the
previous section, the confidence\ region
estimate $I'(y)$ at $\theta=0$ has to be modified in order to take
advantage of the alternative model $\theta=1$
(so that $\Omega_0'\cap\Omega_1=\varnothing$).
Such adaptation does not occur in the confidence distribution approach.
As noted by the author during the discussion of the bounded parameter example,
the confidence distribution estimate does not change when we restrict
the model space,
and this phenomenon is rather odd. The author
dismissed this problem as a secondary issue because it does not change
the semantics of the confidence region
in the frequentist interpretation. However, if we are interested in
achieving (near) optimality for the estimated confidence region, then
this issue becomes a more serious
concern because it means that this simple method ignores a significant
amount of available information
that could have been used in more complicated algorithms.
In conclusion, while the confidence distribution approach is simple to
apply, the simplicity is achieved by
ignoring some useful information.
Therefore, we have to keep the limitations of this method in mind
whenever it is applied to complex statistical models.\vspace*{-2pt}

\end{document}